# Elemental substitution tuned magneto-elastoviscous behavior of nanoscale ferrite MFe$_2$O$_4$ (M = Mn, Fe, Co, Ni) based complex fluids

**Ankur Chattopadhyay** [a,†], **Subhajyoti Samanta**[b], **Rajendra Srivastava**[b], **Rajib Mondal**[c] and **Purbarun Dhar** [a,*]

[a] Department of Mechanical Engineering, Indian Institute of Technology Ropar,

Rupnagar–140001, India

[b] Department of Chemistry, Indian Institute of Technology Ropar,

Rupnagar–140001, India

[c] Department of Condensed Matter Physics and Material Science,

Tata Institute of Fundamental Research, Mumbai–400005, India

[*] _Corresponding author_:

E–mail: purbarun@iitrpr.ac.in

Phone: +91–1881–24–2173

[†] E–mail: ankur.chattopadhyay@iitrpr.ac.in

# Abstract

The present article reports the governing influence of substituting the $M^{2+}$ site in nanoscale $MFe_2O_4$ spinel ferrites by different magnetic metals (Fe/Mn/Co/Ni) on magnetorheological and magneto-elastoviscous behaviors of the corresponding magnetorheological fluids (MRFs). Different doped $MFe_2O_4$ nanoparticles have been synthesized using the polyol-assisted hydrothermal method. Detailed steady and oscillatory shear rheology have been performed on the MRFs to determine the magneto-viscoelastic responses The MRFs exhibit shear thinning behavior and augmented yield characteristics under influence of magnetic field. The steady state magnetoviscous behaviors are scaled against the governing Mason number and self-similar response from all the MRFs have been noted. The MRFs conform to an extended Bingham plastic model under field effect. Transient magnetoviscous responses show distinct hysteresis behaviors when the MRFs are exposed to time varying magnetic fields. Oscillatory shear studies using frequency and strain amplitude sweeps exhibit predominant solid like behaviors under field environment. However, the relaxation behaviors and strain amplitude sweep tests of the MRFs reveal that while the fluids show solid-like behaviors under field effect, they cannot be termed as typical elastic fluids. Comparisons show that the $MnFe_2O_4$ MRFs have superior yield performance among all. However, in case of dynamic and oscillatory systems, $CoFe_2O_4$ MRFs show the best performance. The viscoelastic responses of the MRFs are noted to correspond to a three element viscoelastic model. The study may find importance in design and development strategies of nano-MRFs for different applications.

# 1. Introduction

Magnetorheological fluids (MRFs), which are colloidal dispersions of magnetic nanoparticles, can be dubbed as 'smart fluids' due to their magnetic field dependent tunable viscosity. Conventional MR fluids consist of magnetic submicron particles dispersed stably in non-magnetic liquids and these fluids show altered rheological behavior in presence of external stimulus (magnetic field) [1]. Consequently, MR fluids have received considerable attention in different technological applications, such as shock absorbers, brakes, clutches, seismic vibration dampers, control valves, torque transducers, polishing fluids, and have also been studied for biomedical applications like drug delivery, cell separation, diagnostics sensor, etc. [2-5]. Therefore, in-depth understanding of the fundamental rheological behavior of MRFs is essential for design and development of such field-responsive systems.

When the MRFs are exposed to a magnetic field, the dispersed magnetic nanoparticles (MNPs) coalesce to form a fibrillar microstructure. Under the action of shear, these fibrils or chains resist hydrodynamic deformation forces, leading to improved rheological parameters [6]. It has been theorized that attachment and detachment of the sidechains from dense clusters of particles is responsible for the dynamic viscoelastic behaviors of MRFs [6]. Some studies have shown that the dipole interactions among MNPs are primarily responsible for the difference in aggregation behavior [7, 8], which in turn governs rheological features. Measurement techniques like neutron scattering [9, 10] and X-ray scattering [11, 12] have been employed by researchers to reveal information regarding the field induced microstructures of the MRFs. Reports have also shown [13] that the stress relaxation process in a ferrofluid can be attributed to both linear chains and the dense bulk aggregates. Theoretical models have also been put forward to determine the viscoelastic responses of MRFs [14, 15]. These models propose that for highly concentrated MRFs, complexly shaped structures, larger than single chains observed in dilute MRFs, are majorly responsible for the augmented viscoelastic responses under field effect.

Zubarev et al. have suggested a few analytical models to explain the rheological behavior of ferrofluids [16]. To examine the structural configurations of MR suspensions under steady and dynamic flow, molecular dynamics [17, 18] simulations have been performed and the results indicate that the responses of chainlike structures are strongly dependent on the orientation relative to the direction of applied shear flow under field stimulus. This ultimately leads to pronounced anisotropic viscous behavior. Brownian dynamics simulations have also been used to understand magnetoviscous responses when the MRF is exposed to weak and strong dipole interactions with varying magnetic fields [19].

Since the morphology and structure of MNPs influence the magnetoviscous responses of the MRFs, selection of proper magnetic nanocolloids have strong implications on the overall rheological behavior. Binary transition metal oxides (BTMOs) (denoted by $AB_2O_4$, where A, B = Co, Ni, Fe, Mn) have attracted considerable attention in this field due to their strong size and shape dependent magnetic properties [20, 21]. Ferrite nanoparticles (FNPs), which fall into the category of BTMOs, possess typical spinel structures. Here A and B are metallic cations positioned at two different crystallographic sites, the tetrahedral (A site) and the octahedral (B site) [22]. The common examples for ferrites are $MFe_2O_4$ (where M = Mn, Fe, Co, Ni etc.) and many of these nanoparticles demonstrate superparamagnetic (SPM) properties. The superparamagnetism is manifested below a critical size of the nanoparticles of around 30 nm in diameter [22]. By regulating the $M^{2+}$ appropriately, the magnetic configurations of $MFe_2O_4$ can be tuned to obtain high magnetic permeability and electrical resistivity. These nanoscale ferrites have been considered as the potential candidates in high-performance electromagnetic [23, 24] and spintronic devices [25, 26].

If the $M^{2+}$ occupies only the tetrahedral sites, the spinel is termed as direct, whereas when it occupies the octahedral sites only, the spinel is called inverse [27]. Thus, the preferred location of $M^{2+}$ is critical to tune the magnetic behavior. Although significant progress has been made in preparing typical FNPs, there are few generic processes for producing $MFe_2O_4$ nanoparticles with the desired size and acceptable size distribution [28]. Many studies report the various synthesis processes to obtain $MFe_2O_4$ FNPs and their magnetic properties [28-31]. However, despite their excellent magnetic behavior even at the nanoscale, the employability of such FNPs in nanocolloidal MRFs have not been explored. In recent years, a few studies have investigated the magnetorheological characteristics of $CoFe_2O_4$ [32-34], $ZnFe_2O_4$ [35], $MgFe_2O_4$ [36] based suspensions. However, these studies primarily concentrate upon steady state rheological characteristics and the more on the methods of preparing the MRFs than their fluid dynamics.

The present study explores the magnetoviscous and magneto-elastoviscous responses of $MFe_2O_4$ ferrite based nanocolloidal MRFs. The study also focuses on the examination of the role of the dopant metal ion $M^{2+}$ on the overall magnetorheological response of the fluids. The objective of the study is also to understand magneto-viscoelastic behaviors of the colloids by tuning the $M^{2+}$ (Fe/Mn/Co/Ni) crystalline location. The ferrites have been synthesized by chemical routes and have been characterized in detail for their physical structure and properties. Steady and oscillatory rheological measurements, at small and medium shear rates, have been studied and the responses are analyzed to deduce the viscoelastic nature of the colloids. Standard and extended viscoelastic and rheological models have also been used to determine the elastic and viscous responses of the colloids under magnetic field. The present article may find importance in design and development of nano-MRFs.

## 2. Materials and methodologies

### 2.1. Synthesis of nanomaterials

The base chemical ingredients, such as $FeCl_3.6H_2O$, $MnCl_2.2H_2O$, $CoCl_2.4H_2O$, $NiCl_2.6H_2O$, and PVP (polyvinyl pyrrolidone) were procured and used as is (Loba Chime Pvt. Ltd., India). Ethylene glycol and poly-ethylene glycol (MW 400) were procured from Spectrochem, India, and $NH_4F$ from Sigma-Aldrich, India. $MFe_2O_4$ nanoparicles were synthesized using the polyol assisted hydrothermal method [37]. In a typical synthesis, 20 mmol of $FeCl_3.6H_2O$ and 10 mmol $MCl_2.2H_2O$ (M = Mn, Co, Ni) were taken in a beaker containing 200 mL ethylene glycol and stirred for 30 mins. Then 3.2 gm of polyethylene glycol (M.W 400) was added to this solution under stirring condition and further stirred for 30 minutes. After complete dissolution, 80 mmol of $NH_4F$ was added slowly to the solution and allowed to settle for an additional 2 h to allow a viscous solution to form. Finally, the solution was transferred to a Teflon lined autoclave and kept at 180 °C for 24 h. After the hydrothermal treatment, the autoclave was cooled to room temperature, and the batch was filtered, washed several times with deionized water and dried overnight (10-12 h) at 80 °C which results in the formation of a black powder of $MFe_2O_4$.

## 2.2. Nanomaterial characterization

X-ray diffraction (XRD) patterns for the nanomaterials were recorded in the 2θ range of 5°-80° with a scan rate of 2°/min (PAN analytical X'PERT PRO diffractometer, the Netherlands) using Cu Kα radiation (λ=0.1542 nm, 40 KV, 45 mA) (Fig. SF1, Refer Supplementary). Nitrogen adsorption-desorption measurements were performed at -200 °C (Quantachrome Instruments, Autosorb-IQ volumetric adsorption analyzer, USA). The specific surface area of the material was calculated from the adsorption data points obtained at $P/P_0$ between 0.05-0.3 from the Brunauer-Emmett-Teller (BET) equation. Field enhanced scanning electron microscopy (FESEM) measurements (ZEISS Supra) were done to determine the morphology of the materials (figure 1 (a)-(h)).

The crystallinity, phase purity, and successful formation of all the ferrite samples were confirmed by Powder X-ray (P-XRD) diffraction in the range of 2θ (5°-80°). All the samples exhibit reflections at 2θ = 30.27°, 35.46°, 43.25°, 53.62°, 57.26°, and 62.97°, which indicates FCC framework structure. Except $NiFe_2O_4$, none of the samples exhibit additional reflection other than the FCC framework, and this eliminates the possibility of presence of any other phases (such as $CoO$, $MnO_2$, $Fe_2O_3$, etc. in $COFe_2O_4$, and $MnFe_2O_4$). In the case of $NiFe_2O_4$, low intensity peaks at 2θ = 48.09° and 51.56° (in addition to the standard FCC of $NiFe_2O_4$) are present due to the formation of impurity phase of $NiO/Fe_2O_3$ [38]. The XRD reflection patterns of spine slightly vary from one another due to the difference in their crystal field stabilization energy in their respective coordination geometry.

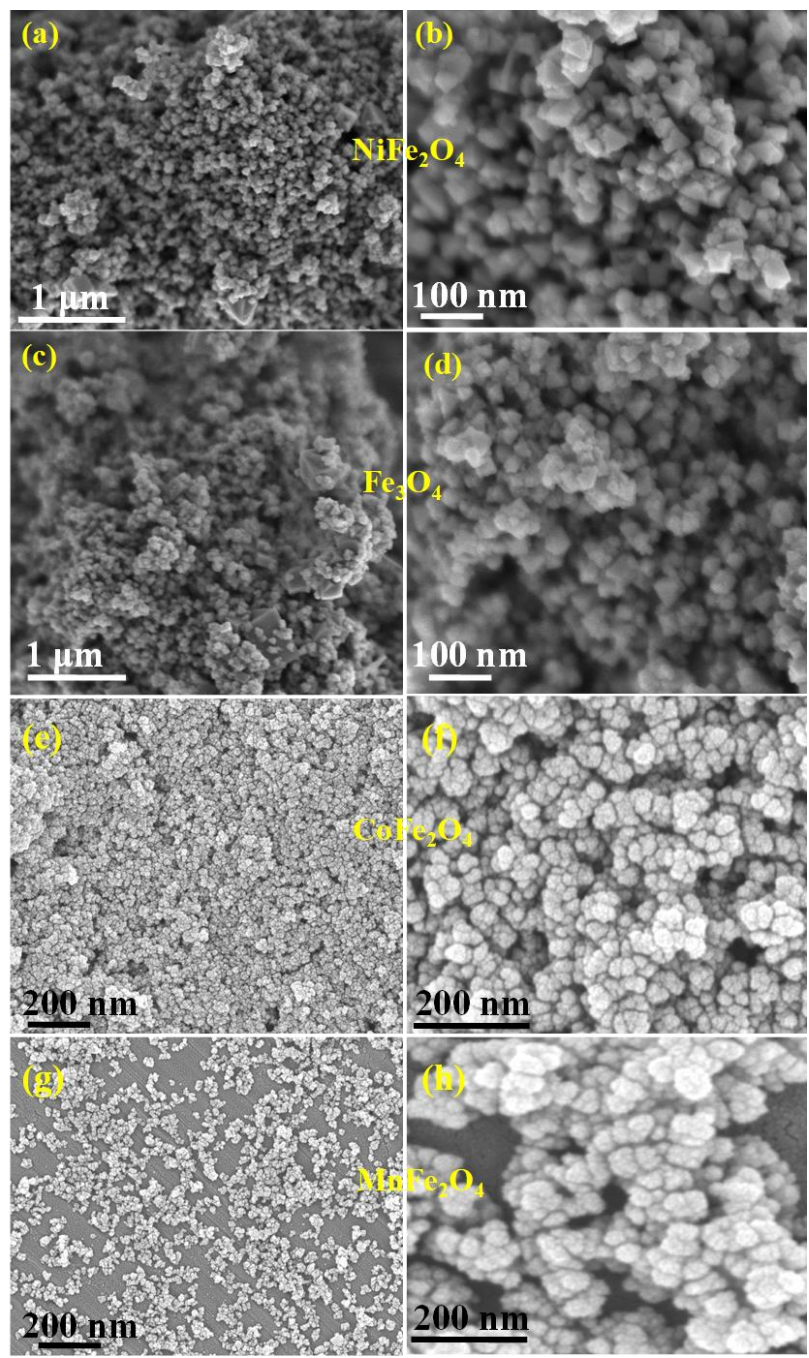

**Figure 1.** FESEM images of (a-b) $NiFe_2O_4$, (c-d) $Fe_3O_4$, (e-f) $CoFe_2O_4$, and (g-h) $MnFe_2O_4$ nanoparticles.

The detailed information regarding crystallite size, FWHM, and other parameters are enlisted in table ST1 (Refer Supplementary). The average crystallite size obtained from P-XRD analysis for $MnFe_2O_4$, $CoFe_2O4$, $NiFe_2O_4$ and $Fe_3O4$ are 10.7, 7.7, 10.4, and 9.2 nm, respectively. The presence and amount of all the elements in a representative material $CoFe_2O_4$ is confirmed from EDAX analysis (Fig. SF2, Refer Supplementary). The surface area obtained from BET measurements and textural properties are summarized in ST 2 (Refer Supplementary). BET analysis further depicts that $CoFe_2O_4$ exhibits highest surface area among the different materials synthesized. Fig. 2 illustrates the room temperature (300K) magnetization curves (M-H) of different $MFe_2O_4$ nanoparticles, measured by Vibrating Sample Magnetometer (VSM). None of the MNPs possess magnetic hysteresis, which confirms superparamagnetic behavior. The Mn based ferrite exhibits the highest saturation magnetization $M_S$ ~ 74 emu /g (achieved within ~ 0.8 T). The saturation magnetization of the ferrites are ~ 62, 44 and 37 emu/g for $Fe_3O_4$, $CoFe_2O_4$, and $NiFe_2O_4$, respectively.

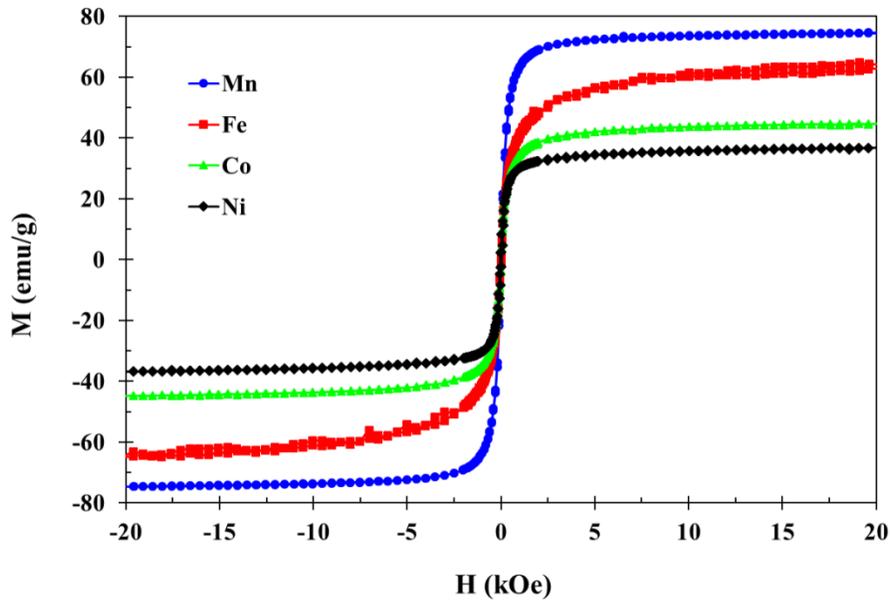

**Figure 2.** Magnetization curves of various $MFe_2O_4$ ferrites at 300K. The absence of magnetization hysteresis confirms the superparamagnetic phase.

## 2. 3. Instrumentation

A non-polar liquid, silicone oil (SO) (procured from Avra Synthesis Ltd., India) was used to prepare the MRFs. The SO has a Newtonian viscosity of 350 cSt at 25 °C. Anhydrous FNPs were dispersed in the SO as per concentration requirements (40 wt. % in the present studies) and stirred mechanically and ultrasonicated to obtain a homogeneous colloid. To prevent moisture adsorption, the MRFs were stored in a desiccator. The MR properties of the colloids were measured using a rotational rheometer (MCR 102, Anton Paar, Germany) with parallel plate configuration at constant gap of 1 mm. The rheometer is connected to a Magnetorheological module, capable of generating magnetic fields up to 1 T for sample thicknesses of 1 mm and lower. The MRFs are tested at four different magnetic field intensities (0, 0.35, 0.7, and 1 T). During the experiments, the sample temperature has been maintained constant at 300K using a Peltier controller. The rheological behavior is obtained by measuring the shear stress and viscosity as a function of shear rate from 0.01 to 100 $s^{-1}$. The viscoelastic responses have been studied in terms of storage modulus (G'), loss modulus (G''), dissipation (loss) factor (tan$\delta$), and the complex viscosity ($\eta^*$). To probe the dynamic rheological response, oscillatory tests with strain amplitudes of 0.01 to 1 % and frequencies of 1 to 100 Hz have been performed. Measurements of stress relaxation behaviors and magnetoviscous hysteresis have also been performed. The typical uncertainty involved in the measurements was within ±5%.

## 3. Results and discussion

### 3.1. Magnetorheology

Figures SF4-SF7 (Refer supplementary) illustrates the role of magnetic field in modifying the static rheological behaviors of the colloids. The increase in viscosity is due to the formation of chain-like structures by the nanoparticles within the colloids under the action of magnetic field. At lower magnetic field strengths, the chains or fibrils formed within the MRF are typically linear. With increase of field strength, the number densities of chains improve with increasing aspect ratio of the fibril structures [39]. Further increase of field strength can lead to lateral coalescence of the chains, resulting in thick columns [40, 41], and ultimately decrease in the magnetic field induced viscosity. All the samples (MRFs) exhibit shear thinning behaviors for the entire range of shear rates and magnetic field strengths. Reports have shown that the chains under confinement can show different responses compared to unconfined chains [42], which is possibly the reason why ferrites with higher saturation magnetization show low viscous response to magnetic fields compared to the lower magnetization ferrites. The higher magnetization leads to agglomeration of the chains at lower field strengths, and the viscous response of the colloid does not improve at higher fields.

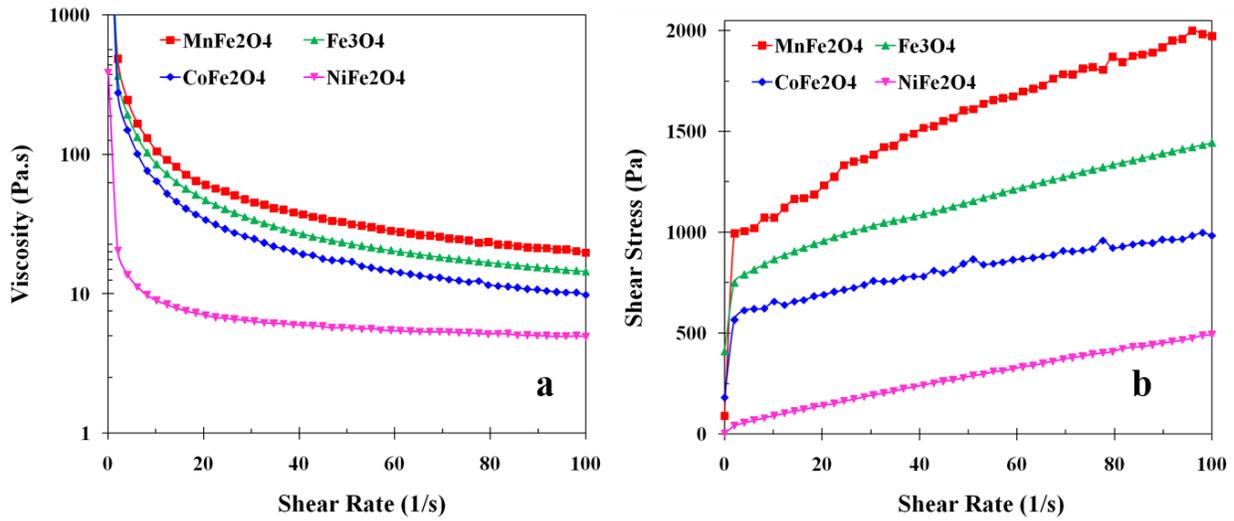

**Figure 3.** (a) Viscosity behavior against shear rate for various FNPs based MRFs at 1T (b) Shear stress response to shear rates for various FNPs based MRFs at 1T.

The Mn-based ferrite colloids show the highest viscous response, followed by the $Fe_3O_4$, which is followed by the Co, Cu and Ni-based ferrites. This can be explained using the coupling constant ($\lambda$). For a constant magnetic field and temperature,

$$\lambda \propto \chi^2 \tag{1}$$

where, $\chi$ is the magnetic susceptibility of the material [43]. Higher the value of $\lambda$ more is the likelihood of formation of chains that are aligned along the direction of the magnetic field. This directly results in enhanced magneto-static particle interaction, leading to enhanced viscosity [42]. The coupling constant is known to have a quadratic relationship with the magnetic susceptibility. Thus, the slope of the M-H curve has a strong effect on the viscosity of the associated MRFs. The magnitudes of $\chi$ for the FNPs follow the order as Mn>Fe>Co>Ni (fig. 2) and consequently, the magnetoviscous effect of their colloids obey the same order (fig. 3 (a)).

The yield stresses of the MRFs also reveal similar magnetorheological behaviors (fig. 3 (b)). The yield stress provides an estimate of the force required to continuously deform the particle aggregates and chains which tend to reform in the presence of the magneto-static forces due to the applied magnetic field. Enhancement of yield values has been observed when the MRFs are exposed to increased magnetic field intensities (Figs. SF4-SF7). The yield stress is known to increase in a quadratic manner with increasing magnetic field strengths at low field regimes as

$$\tau_y \propto \mu_0 \mu_c (\beta H)^2 \tag{2}$$

where $\mu_0$ is the permeability of the vaccum, $\mu_c$ is the relative permeability of the carrier fluid. In eqn. 2, $\beta$ is the contrast factor, defined as

$$\beta = \frac{\mu_p - \mu_c}{\mu_p + 2\mu_c} \tag{3}$$

where $\mu_p$ is the relative permeability of the particle. For instance the $\beta$ of $Fe_3O_4$ based MRF comes around 2.7 [44]. The respective values of $\beta$ of other MRFs vary from 2 to 3, based on the choice of nanomaterials.

The augmentation of yield stress is pronounced in the linear regime of magnetization [45 - 47]. At higher magnetic fields, around the magnetic saturation limit, the yield stress becomes field independent

$$\tau_y \propto \mu_0 M_s^2 \tag{4}$$

where $M_s$ represents the saturation magnetization. It is important to understand the comparative yield stresses of the MRFs (fig. 4 (a)) caused due to substitution of Fe by other metals (Mn, Co, Ni etc.) in the ferrites. Doping with Mn in $MFe_2O_4$ results in superior yield stress compared to Fe. However, doping with Co and Ni leads to reduction in yield stresses than $Fe_3O_4$ MRFs. These observations signify that there is a strong dependence of the yield stress on the magnetic moment of the $M^{2+}$ (n$\mu_B$, where n = 5, 4, 3 and 2, for Mn, Fe, Co, and Ni, respectively) [29, 48, 49]. Hence, the static magnetorheological behavior in ferrite based MRFs is a strong function of the magnetic properties of the dopant atom.

Fig. 4 (a) illustrates the yield stress values for three values of applied magnetic fields. Typically, the rate of enhancement of yield stress values is high for low magnetic field strengths, whereas, at high field strengths, the yield stress values reach saturation. An extended Bingham model has been used to predict the yield stress at both low and high regimes of magnetic field strength. It is expressed as [50]

$$\tau_y = \tau_{y\infty} + 2(\tau_{y0} - \tau_{y\infty})(e^{-\alpha H} - \frac{e^{-2\alpha H}}{2}) \qquad (5)$$

where, $\tau_y$ is the yield stress at magnetic field $H$. In eqn. 4, $\tau_{y0}$ represents the yield stress at zero-field, and $\tau_{y\infty}$ represents the corresponding yield stress once it has attained saturation. The parameter $\alpha$ is a fit variable in the extended Bingham model, which have been shown in table 1. The Mn-based MRF has the highest saturation yield stress and hence for a particular imposed field, Mn-based MRFs can resist higher deformation.

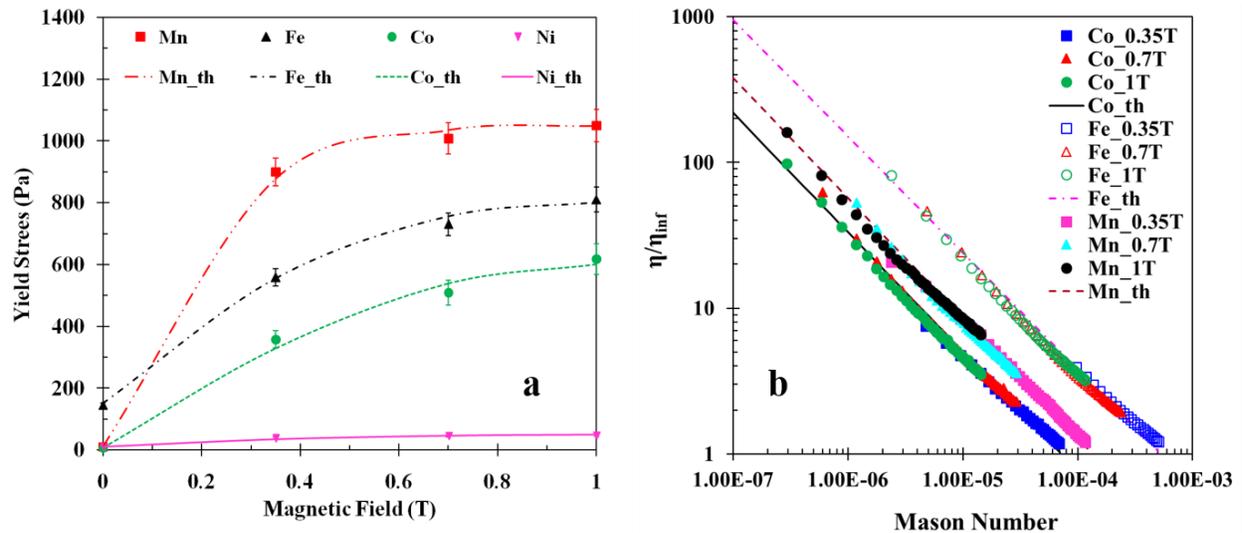

**Figure 4.** (a) Comparison of experimental yield stress values and the model predictions (eqn. 4) [symbols refer to experimental observations and lines represent the model], (b) comparison of viscous behaviors of different ferrite (Fe/Mn/Co) MRFs as function of Mason number

**Table 1.** Fit parameters of yield stress model for different MRFs

| Parameters | Mn | Fe | Co | Ni |
|---|---|---|---|---|
| $\tau_{y\infty}(Pa)$ | 1050 | 820 | 635 | 50 |
| $\tau_{y0}(Pa)$ | 10 | 150 | 6.5 | 10 |
| $\alpha$ | 7 | 4.3 | 3.6 | 4.2 |

As observable in fig.4, the present MRFs conform to the Bingham plastic model, expressible as

$$\tau = \tau_y + \eta_{pl}\dot{\gamma} \quad (6)$$

where, the shear stress ($\tau$) is related to the yield stress ($\tau_y$), the plastic viscosity ($\eta_{pl}$) and the imposed shear rate ($\dot{\gamma}$). The apparent viscosity $\eta_{app}$ for the MRF is defined as

$$\eta_{app} = \frac{\tau}{\dot{\gamma}} \quad (7)$$

Introducing the non-linear nature of the plastic viscosity, the apparent viscosity can be remodeled in terms of the infinite shear viscosity as

$$\frac{\eta_{app}}{\eta_\infty} = \frac{\tau_y}{\eta_\infty \dot{\gamma}} + \frac{\eta_{pl}}{\eta_\infty} \quad (8)$$

It is shown by reports that $\eta_{pl}$ would ultimately be equal to $\eta_\infty$, especially at high shear limits [51, 52]. The non-dimensionalized form of the eqn. 6 can be expressed as

$$\frac{\eta_{app}}{\eta_\infty} = \frac{\psi}{Mn} + 1 \quad (9)$$

where, $\psi$ is a constant which is determined from fitting the experimental observations to the eqn. 9. The Mason number ($Mn$) is the ratio of the hydrodynamic shear forces to magnetic (ferrodynamic) forces in MRFs and is expressed as

$$Mn = \frac{\eta_c \dot{\gamma}}{2\mu_0 \mu_c (\beta H)^2} \tag{10}$$

Fig. 4 (b) illustrates the viscosity ratio as a function of Mason number by adjusting the set of $\psi$ and $\eta_\infty$ to obtain the best possible fit. In case of the present MRFs, $\psi$ is deduced to range between $4 \times 10^{-4}$ and $2 \times 10^{-3}$ and $\eta_\infty$ ranges in ~ 7-12 Pas for the entire range of magnetic field strengths.

Understanding the transient responses of MRFs is important to assess their performances in dynamic environments. Figure 5 (a) illustrates the magnetic field sweep test results for $Fe_3O_4$ MRFs at different shear rates (0.01 s$^{-1}$ to 1 s$^{-1}$). The viscosity enhances with increase in magnetic field and saturates at high fields, irrespective of the magnitude of the shear rates. The magnetic field leads to formation of the chained fibrils by the aligned nanoparticles within the MRFs. Such structure induces localized elasticity to the fluid phase and leads to enhanced shear resistance, or viscosity. The number density of the fibrils reach a maximum at a particular field, and beyond that, no additional fibril formation takes place. Thereby, the magnetoviscous behavior reaches a plateau. The MRFs show hysteresis in the magnetoviscous effect (fig. 5 (a)) upon application and withdrawal of magnetic field. It can be seen in fig. 5 (a) that the curves for the decreasing field case lies above the increasing field case. This suggests that the field-induced chains or fibrils possess a structural hysteresis [53], even though the constituent nanoparticles are superparamagnetic. With the withdrawal of the field, the magnetization of the nanoparticles relax immediately (due to zero magnetic hysteresis), however, the interparticle magneto-static interactions require a finite relaxation period. This causes the elasticity of the microstructure to relax in a lagging manner to the external field, leading to the positive magnetoviscous hysteresis at lower fields. The comparative performances of all the ferrite MRFs for a shear rate of 1 s$^{-1}$ have been illustrated in fig. 5 (b). It is noteworthy that all the doped ferrite MRFs show reduced magnetoviscous hysteresis compared to the $Fe_3O_4$ MRFs.

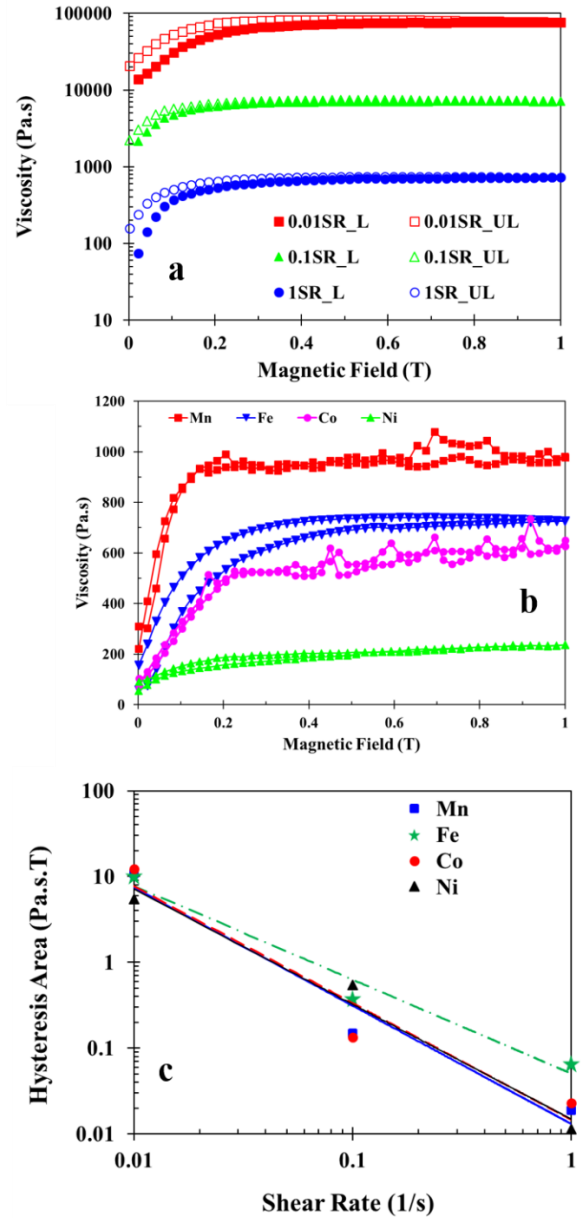

**Figure 5.** (a) Comparison of magnetic hysteresis of $Fe_3O_4$ MRFs for different shear rates (0.01, 0.1, 1 s$^{-1}$) ('L' represents loading – increasing magnetic field – *solid symbol*, and 'UL' represents unloading - decreasing magnetic field - *open symbol*), (b) comparison of magnetic hysteresis of $MFe_2O_4$ (M = Mn, Fe, Co, Ni ) MRFs at shear rate of 1 s$^{-1}$, (c) comparison of magnetoviscous hysteresis area between experimental data and theoretical model of Fe/Mn/Co/Ni ferrite based MRFs [symbols represent experimental observations and lines represent eqn. 13]

Upon application or withdrawal of the magnetic field, the magnetic moment of a particle can relax by two mechanisms. In case of the Brownian relaxation, the magnetic moment remains fixed within the particle, and it reorients its position as a whole. The rotation of magnetic moment instead of the actual particle conformation is called the Neel relaxation. It is common for smaller particles (size of the order ~ nm) to magnetically relax by Neel mechanism, while larger particles tend to follow the Brownian mechanism. The Brownian relaxation time ($\tau_B$) and Neel relaxation time ($\tau_N$) scales can be expressed as per eqns. 11 and 12 respectively.

$$\tau_B = \frac{3\tilde{V}\eta_c}{k_B T} \tag{11}$$

$$\tau_N = \left(\frac{1}{f_0}\right)\exp\left(\frac{KV}{k_B T}\right) \tag{12}$$

where, $f_0$ is the frequency of domain flipping, whose value has been assumed to be $10^9$ Hz for magnetic particles in non-polar media [54], $K$ is the magnetic anisotropy constant, $k_B$ is the Boltzmann constant, $T$ represents the absolute temperature and $V$ is the volume of the FNP. The typical values of $\tau_B$ and $\tau_N$ of the FNPs in the present study are ~ $10^{-6}$ s and ~ $10^{-9}$ s, respectively. The process with the smaller relaxation time governs the overall relaxation of the MRFs [54], and in the present case, the Neel relaxation is the dominant mechanism.

It can be observed from fig. 5 (a) that magnetoviscous remanence is a function of the imposed shear rates. The area enclosed by the loading (increasing field) and unloading (decreasing field) curves, referred to as the magnetoviscous hysteresis area, can be used to model such MRFs. When the hysteresis areas are plotted as function of shear rate, relative decrease are observable with increasing shear values (fig. 5 (c)). This behavior can be predicted using power law model as per [42]

$$A = u\dot{\gamma}^v \tag{13}$$

where, $A$ is the predicted magnetoviscous hysteresis area, $u$ is a scaling factor, and $v$ is an exponent of the shear rate ($\dot{\gamma}$). The respective values of $u$ and $v$ of different MRFs have been listed in Table 2. The reduction of hysteresis areas at increased shear rates indicates that the magnetic moment remanence due to interparticle interactions within the fibrils are overcome by the higher shear, which leads to loss of structural integrity of the fibrils. This leads to quicker lowering of the viscosity on withdrawal of the field, leading to lower viscous hysteresis. A point of interest may be noted from table 2. For the doped ferrite based MRFs, the values of $u$ and $v$ are very similar, which suggests that these fluids exhibit self-similar magnetoviscous hysteresis. This point may find useful implications in dopant based magnetic nanoparticle synthesis for control of magnetoviscous hysteresis in the corresponding MRFs.

**Table 2.** Fit parameters of eqn. 13 of different ferrite MRFs (goodness of fit >0.95)

| *Parameters* | **Mn** | **Fe** | **Co** | **Ni** |
|---|---|---|---|---|
| *u* | *13* | *50* | *15* | *15* |
| *v* | *-1.38* | *-1.1* | *-1.36* | *-1.34* |

### 3.2. Magneto-viscoelasticity

Oscillatory rheological responses of the MRFs have been examined to understand their dynamic behaviors. Fig. 6 (a) illustrates the magneto-viscoelastic behavior of the MRFs. The storage (G') and loss (G") moduli for the different MRFs are obtained from frequency sweep tests at different magnetic field strengths at constant strain amplitudes. In absence of magnetic field, for the entire range of frequencies, the G" is higher than the G', signifying predominantly liquid behavior. Under the influence of magnetic field, the improvement of G' component indicates microstructure elasticity within the MRF. The particles within an MRF experience rotational resistance when in the presence of a magnetic field, due to competition between magnetic and hydrodynamic torques [55, 56]. The hydrodynamic torque tends to distort the alignment of particles induced by the magnetic field. At higher fields, the hydrodynamic torque on the fibrils is overcome by the magnetic torque, thus leading to a microstructure with elastic integrity. This leads to a behavior which mimics polymeric or elastic fluids (typically linear viscoelasticity behavior at high fields (fig. 6 (a)). The loss factors also exhibit improvement under field stimuli (fig. 6 (b)), which indicates enhanced dissipative behavior in addition to improved elasticity and is a preferred characteristics for dynamic applications. With increase in frequency, the adjacent chains are repeatedly sheared in each other's vicinity. Consequently, the hydrodynamic torque exceeds the magnetic torque, leading to viscous behavior (loss factor approaches unity) at high frequencies.

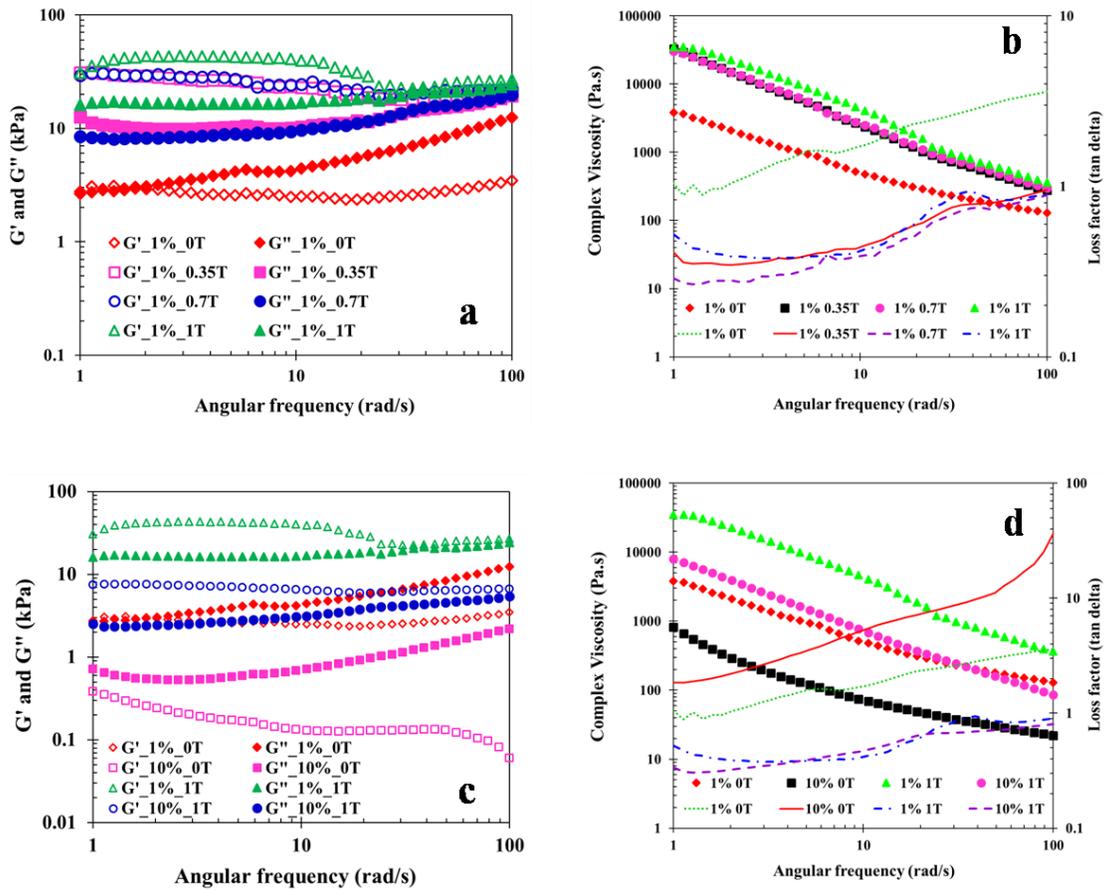

**Figure 6.** Frequency sweep responses at different magnetic field strengths (0, 0.35, 0.7, 1 T) (a) storage and loss moduli of $Fe_3O_4$ MRF at 1% strain amplitude [open symbol – storage modulus, closed symbols – loss modulus], (b) complex viscosity and loss factors as function of oscillatory frequency 1% strain amplitude [symbol – complex viscosity, line – loss factor], (c) storage and loss moduli at 10% strain amplitude [open symbol – storage modulus, closed symbols – loss

modulus], (d) Complex viscosity and loss factor at 10% strain amplitude [symbol – complex viscosity, line – loss factor]

The comparisons between frequency sweep responses at strain amplitudes of 1% and 10% for $Fe_3O_4$ MRFs have been illustrated in figs. 6 (c) and 6 (d). The behaviors of 10 % are similar in nature to the 1 % case, which signifies good phase stability at higher oscillatory frequency values. Scrutiny of fig. 6 (d) reveals that while the 10 % case shows higher lossy behavior than the 1 % case at 0 T, the loss factors are similar at 1 T. This signifies that the MRFs possess higher microstructural elasticity at higher frequencies and high fields. This behavior could be of potential interest for usage in utilities. Fig. 7 illustrates the role of elemental substitution on the frequency sweep viscoelastic behavior of the different MRFs. It is noteworthy that the doping by the Mn does not lead to improvement in the magneto-viscoelastic response of the MRFs compared to the $Fe_3O_4$ based MRFs. While in the steady shear rheology the role of magnetization moment of the doping element was prominent, the dynamic or oscillatory rheology case is not so.

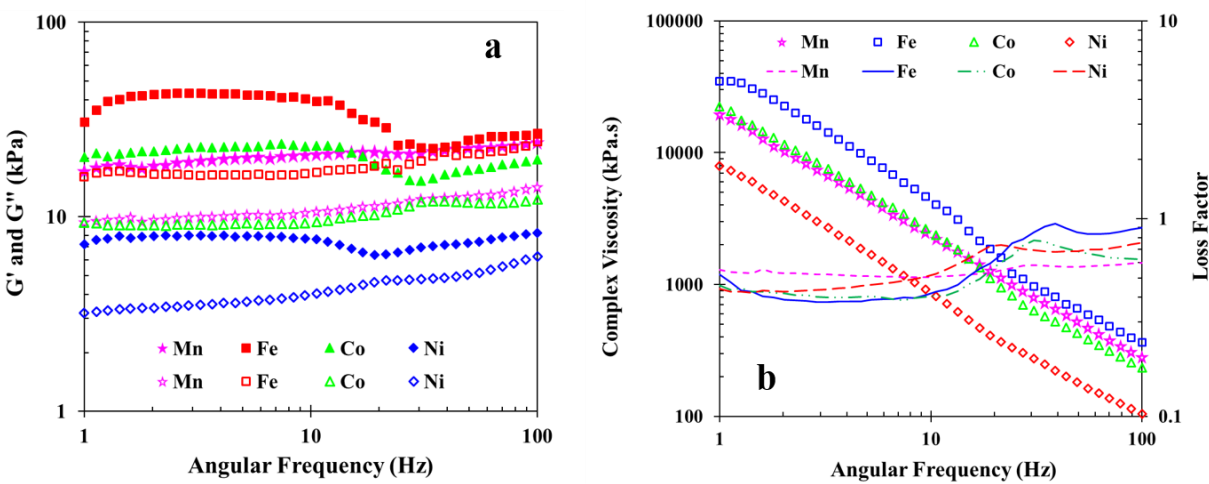

**Figure 7.** Frequency sweep responses of various $MFe_2O_4$ MRFs at 1 T at strain amplitude 1% (a) storage and loss moduli [open symbol – storage modulus, solid symbols – loss modulus], (b) complex viscosity and loss factor as functions of oscillatory frequency [symbol – complex viscosity, line – loss factor]

To understand the role of imposed oscillatory strain on the viscoelastic behavior of the MRFs, amplitude sweep experiments (from strains of 0.01% to 1%) are performed for different oscillatory frequencies. Figs. 8 (a) and 8 (b) illustrate the amplitude sweep viscoelastic responses of $Fe_3O_4$ MRFs at oscillatory frequencies of 1 and 10 Hz. The comparison of the behaviors of the doped ferrite MRFs have been illustrated in figs. 8 (c) and 8 (d). The G" is higher than the G' at zero field case (figs. 8 (a) and (b)) with no distinct region of linear viscoelasticity. With magnetic field, the G' is higher than the G" at low values of strain amplitude. As the strain amplitude increases, a distinct crossover is noted and the viscous behavior overshoots the elastic behavior.

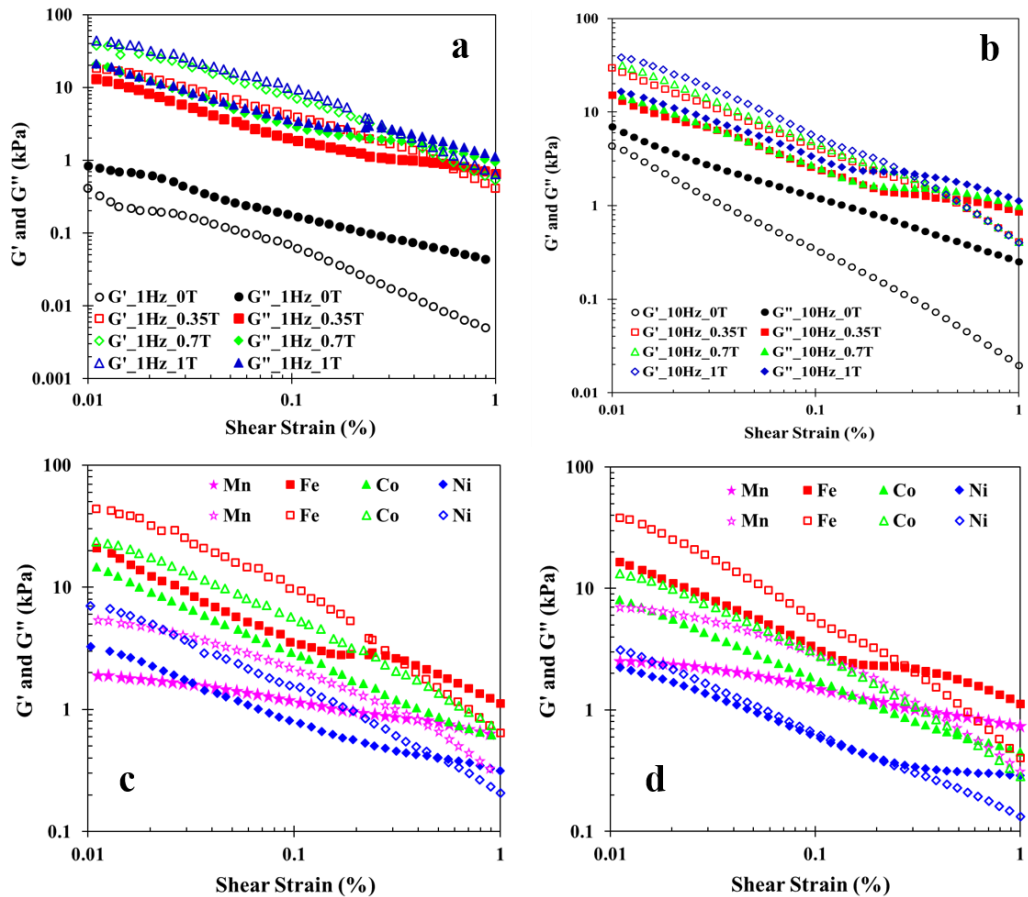

**Figure 8.** Amplitude sweep responses (a) storage and loss moduli of $Fe_3O_4$ MRFs at different magnetic fields (0, 0.35, 0.7, 1 T) at 1 Hz oscillatory frequency [open symbol – storage modulus,

closed symbols – loss modulus], (b) storage and loss moduli at 10 Hz [open symbol – storage modulus, closed symbols – loss modulus], (c) comparison of various MRFs at 1 Hz and 1T [open symbol – storage modulus, closed symbols – loss modulus], (d) comparison of various MRFs at 10 Hz and 1T [open symbol – storage modulus, closed symbols – loss modulus].

Under the effect of magnetic field also no linear regime is observed. The crossover amplitudes have been illustrated in fig. 9 (a). Increase in the field strength increases the crossover amplitude, thereby postponing the initiation of viscous deformation. Additionally, higher oscillatory frequencies lead to reduction in the crossover amplitude due to lesser relaxation time available to the fibrils to align with respect to the magnetic field. This leads to loss of microstructure elasticity locally, and leads to reduction in crossover amplitude. It is noteworthy that the MRFs of ferrites of higher magnetic moments (such as Mn based ferrite and $Fe_3O_4$) exhibit reduction in the crossover amplitude at higher magnetic field strengths. This can be explained based on the inter-chain or inter-fibril magnetic interactions in such MRFs. Due to the high magnetic moment of the constituent particles in the chains, the magneto-static repulsion between neighboring chains will be high. This leads to decrease of the effective elasticity of the microstructure, and even more so at higher fields. Consequently, the crossover amplitude of the MRFs of such ferrites deteriorates at high fields.

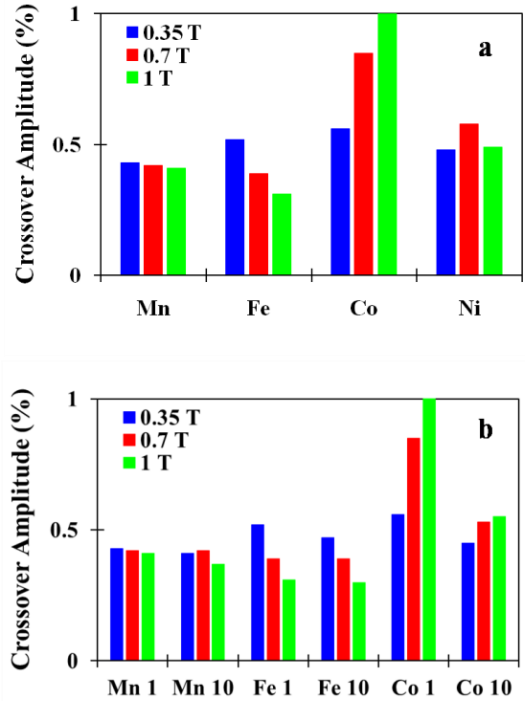

**Figure 9.** Crossover amplitude (a) comparison of various MRFs at 1 Hz for various magnetic field strengths (0.35, 0.7, 1 T), (b) comparison for different MRFs at 1 Hz and 10 Hz.

The relaxation behaviors of the MRFs have also been characterized to determine their viscoelastic nature. Figs. 10 (a) and (b) illustrate the stress relaxation behaviors of the $Fe_3O_4$ and $MnFe_2O_4$ MRFs respectively for a given magnitude of 10% strain. The relaxation modulus improves with increase in field strength. At higher strains, the nature of the transient evolution of the relaxation curve is similar for $MnFe_2O_4$ MRFs, however in case of $Fe_3O_4$ MRFs, the transient evolution changes with field strength. The presence of the field induces fibrillation within the MRF. The fibrils lead to local elasticity within the fluid. The viscoelastic nature leads to increased relaxation modulus as well as weaker temporal relaxation (fig. 10 (b)), implying that the microstructure withstands the applied strain without plastic deformation to a greater extent. To quantify the magnitude of stress relaxation, a relaxation ratio has been defined and the values have been illustrated (figs. 10 (c) and 10 (d)). It is defined as the ratio of the initial relaxation modulus to the relaxation modulus (at 100s in the present case). It is observable from fig. 10 (c) that Mn based MRFs exhibit the highest stress relaxation caliber under field constraints.

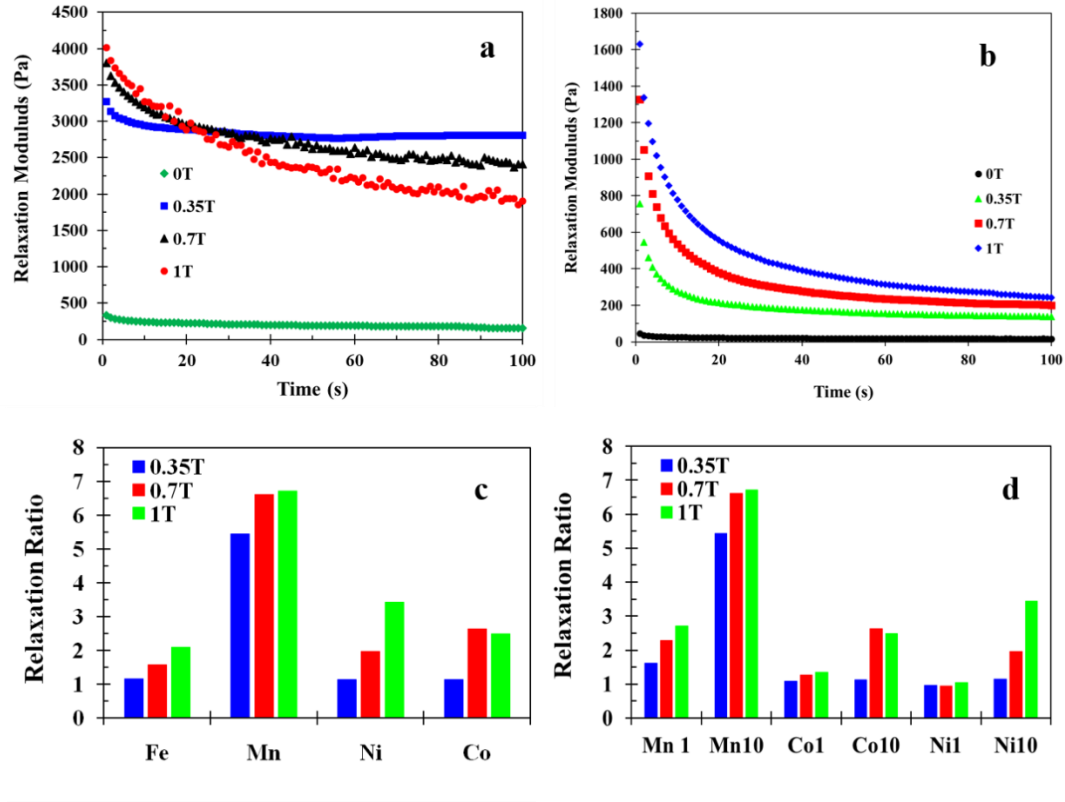

**Figure 10.** (a) relaxation modulus of $Fe_3O_4$ MRFs at 10% strain, (b) relaxation modulus of $MnFe_2O_4$ MRFs at 10% strain, (c) relaxation ratios of various MRFs at 10 % strain, (d) comparison of relaxation ratios of different MRFs at 1% and 10% strains.

The viscoelastic responses of the MRFs have been compared against theoretical models to determine the type of elastic fluids that the present fluids conform to. Typically, viscoelastic materials can be modeled as different combinations of elastic springs (signifies conservation of energy) and viscous dashpots (represents dissipation of energy), which ultimately leads to different constitutive stress–strain relations [57]. A simple viscoelastic system can be modeled as two element classical models, like Maxwell model (fluid like behavior) or Kelvin-Voigt model (solid like behavior) [58, 59]. The present MRFs, especially under magnetic field, do not conform to the two element models. Accordingly, 3 element elastic and viscous models have been implemented to model the experimental observations. It has been found that a three element elastic model also does not hold good for the MRFs, and hence the three element viscous model has been employed. The model consists of a viscous dashpot, which is connected in series with a classical Kelvin-Voigt element (a spring and dashpot connected in parallel). The schematic of the model element has been shown in fig. SF8 (Refer Supplementary). For a Newtonian fluid based damping element, the stress within the dashpot element is expressed as

$$\sigma = \eta \frac{d\varepsilon}{dt} \tag{14}$$

where, $\sigma$ is the stress, $\varepsilon$ is the strain and $\eta$ is the viscosity of the fluid.

However, such a Newtonian three element fluid model fails to replicate the experimental viscoelastic moduli. This is atypical for the present case, as prominent linear viscoelastic regimes were not identifiable from the amplitude and frequency sweep studies. A fractional order time derivative $q$ is introduced in the constitutive equation of the dashpot to provide a realistic representation of a non-linear, non-Newtonian fluid system. For a non-linear, non-Newtonian dashpot, the stress is expressed as

$$\sigma(t') = k \frac{d^q \varepsilon}{dt^q} = k \left[ \frac{1}{\tau^q} \frac{d^q \varepsilon}{d(t/\tau)^q} \right] = \beta \frac{d^q \varepsilon}{dt'^q} \tag{15}$$

where, $\beta = k\tau^{-q}$ is a variable equivalent to viscosity and governs the viscous relaxation behavior, $\tau$ is the characteristic time and $t'$ is the dimensionless time [60-62]. The net stress within the Voigt element of the 3 element systems is [63]

$$\sigma(t') = E\varepsilon_2 + \beta_2 \frac{d^q \varepsilon_2}{dt'^q} \tag{16}$$

where, E is the elastic modulus of the associated spring component within the Voigt element.

The stress within the dashpot which is placed in series with the Voigt element is as (the index $p$ governs the fraction response of the series dashpot)

$$\sigma(t') = \beta_1 \frac{d^p \varepsilon_1}{dt'^p} \tag{17}$$

The total strain $\varepsilon$ within the 3 element system is determined via Laplace transform as

$$\varepsilon(s) = \frac{\sigma(s)}{\beta_1 s^p} + \frac{\sigma(s)}{E + \beta_2 s^q} \tag{18}$$

The final form of the stress-strain relationship in the frequency domain is [63]

$$\sigma(\omega) = \varepsilon(\omega) \frac{\beta_1 (i\omega)^p (E + \beta_2 (i\omega)^q)}{E + \beta_2 (i\omega)^q + \beta_1 (i\omega)^p} \tag{19}$$

Expanding the eqn. 19 in terms of the storage and loss moduli yields [63]

$$G'(\omega) = \frac{2\beta_1 \beta_2 E \omega^{p+q} fg + \beta_1^2 E \omega^{2p} + \beta_1 E^2 \omega^p f + \beta_1^2 \beta_2 \omega^{2p+q} g + \beta_1 \beta_2^2 \omega^{p+2q} f}{(E + \beta_2 \omega^q g + \beta_1 \omega^p f)^2 + (\beta_2 \omega^q e + \beta_1 \omega^p d)^2} \tag{20}$$

$$G''(\omega) = \frac{2\beta_1 \beta_2 E \omega^{p+q} dg + \beta_1 E^2 \omega^p d + \beta_1^2 \beta_2 \omega^{2p+q} e + \beta_1 \beta_2^2 \omega^{p+2q} d}{(E + \beta_2 \omega^q g + \beta_1 \omega^p f)^2 + (\beta_2 \omega^q e + \beta_1 \omega^p d)^2} \tag{21}$$

where, $\sin(p\pi/2) = d$, $\sin(q\pi/2) = e$, $\cos(p\pi/2) = f$ and $\cos(q\pi/2) = g$. Value sets of $E$, $\beta_1$, $\beta_2$, $p$ and $q$ are estimated to predict the frequency sweep observations for the MRFs. The Mn based MRF has been used as a representative system and eqns. 20 and 21 have been fit to the viscoelastic moduli (fig. 11). The set values of the different parameters that fit the viscoelastic responses have been given in tables ST3 and ST4 (Refer Supplementary). The values of $p$ and $q$ lie between 0 and 1 for entire range of magnetic field intensities. This justifies the use of fractional viscoelasticity model for the present MRFs. With increase in field strength, the values of $E$, $\beta_1$ and $\beta_2$ are found to enhance.

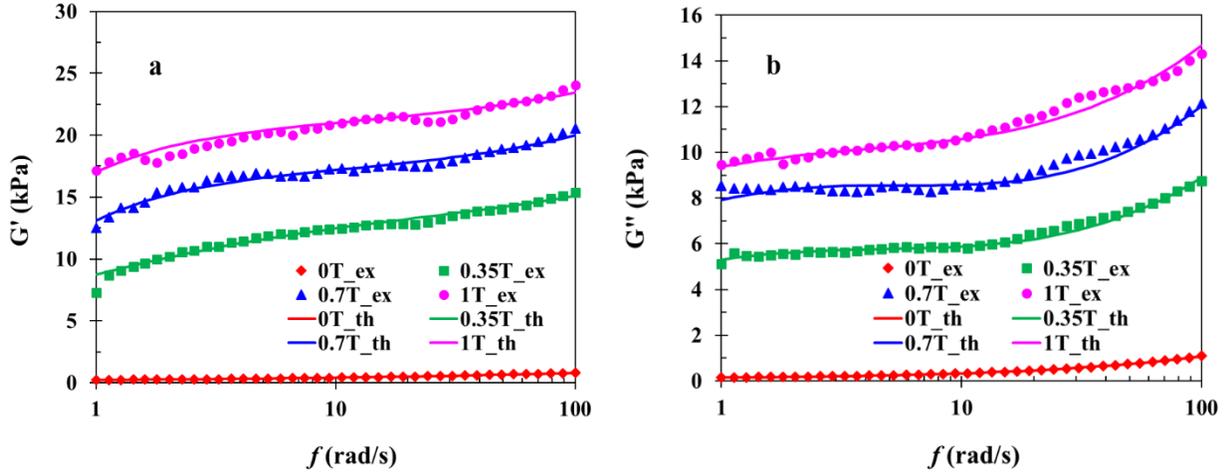

**Figure 11.** Comparison of experimental observations (at 1% strain) for Mn based MRFs with the 3 element fractional fluid model (a) storage modulus and (b) loss modulus. [The symbols represent the experimental observations and the theoretical predictions are indicated by solid lines]

Additionally, the relaxation modulus (G(t)) also can be deduced from eqn. 16 and can be expressed as [64]

$$G(t) = E\left[1 + \frac{(t/t_c)^{-q}}{\Gamma(1-q)}\right] \tag{22}$$

Where, $t_c$ is the time constant and follows the relationship

$$t_c^q = \frac{\beta_2}{E} \tag{23}$$

Substitution of eqn. 23 in the eqn. 22 leads to the simplified form of the relaxation modulus as

$$G(t) = E + \frac{\beta_2 t^{-q}}{\Gamma(1-q)} \tag{24}$$

Fig. 12 illustrates the comparison of the estimated G(t) from the eqn. 24 and the experimental data. The responses of MnFe$_2$O$_4$ and Fe$_3$O$_4$ based MRFs can be distinguished from the respective choices of $E$, $\beta_2$ and $q$; and have been tabulated in tables ST5 and ST6 (Refer Supplementary). This indicates that the viscous dashpot and the elastoviscous Voigt element are play similarly important roles on the viscoelastic responses and temporal relaxation behaviors under magnetic field.

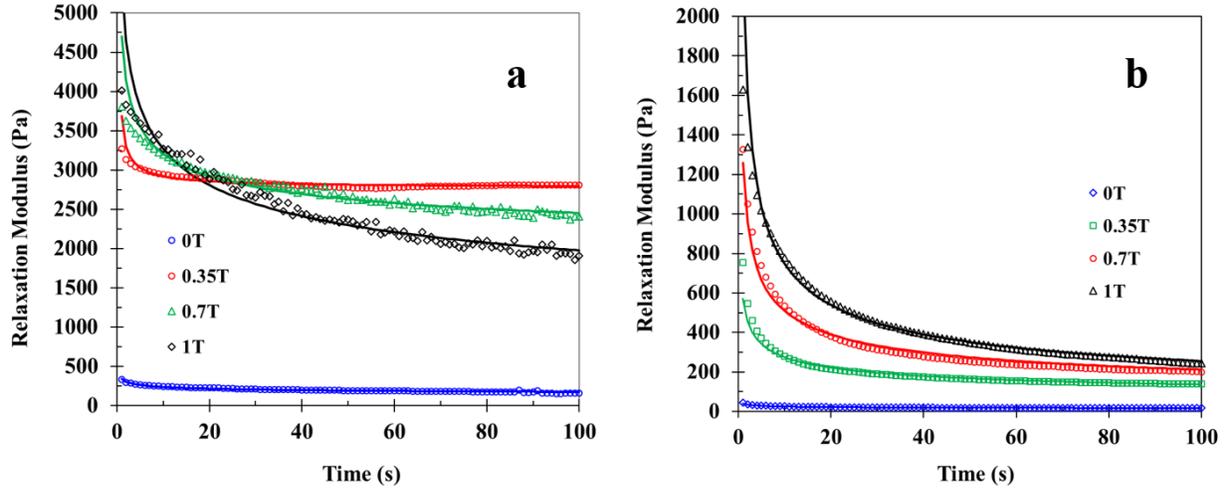

**Figure 12.** Comparison of experimental relaxation moduli (at 10% strain) with the estimations from eqn. 24 (a) $Fe_3O_4$ based MRFs, (b) $MnFe_2O_4$ based MRFs [The symbols represent the experimental observations and the theoretical predictions are indicated by solid lines]

## 4. Conclusions

The present article discusses the magnetoviscous and magneto-elastoviscous behaviors of nanoscale $MFe_2O_4$ ferrites based MRFs. The article is aimed at understanding the role of cation substitution of magnetic ferrites on the magnetic behavior of their MRFs. $MFe_2O_4$ nanoparticles have been synthesized using polyol-assisted hydrothermal method. The detailed characterization confirms the presence of Mn, Co, Ni etc. in $MFe_2O_4$ nanoparticles as the doped atom. The nanoparticles have been dispersed in silicone oil (40 wt. % concentration) and ultrasonicated as required to obtain the corresponding MRFs. The steady shear rheometry reveals that $MnFe_2O_4$ based MRFs show the yield stress among all the other fluids. The order of yield stress improvement under magnetic field is noted to a trend similar to the magnetic moments of the ferrite nanoparticles. The MRFs behave like typical Bingham plastic under magnetic field.

Comparison with an extended Bingham plastic model shows good agreement with the experimental observations for all magnetic field strengths. The magnetorheological behaviors of all the MRFs are self-similar, and all the apparent viscosities under field influence conform to a master curve against the governing Mason numbers. The transient responses of the MRFs show distinct magnetoviscous remanence patterns. The hysteresis areas of the MRFs comply to a proposed simple power law correlation.

Improved viscoelastic responses, in terms of loss factors, have been observed for the different MRFs during magneto-viscoelasticity tests. While in the absence of magnetic field, the MRFs are predominantly fluidic, presence of magnetic field induces elastic nature to the fluids. In the viscoelastic studies also the role of the doped magnetic cation is prominent, and conforms to the nature of magnetization of the cation. Oscillatory strain tests reveal that the MRFs do not possess any distinct linear viscoelastic response, and hence the fluids are not typically elastic fluids by definition despite possessing G' higher than G". But the MRFs show distinct crossover locations and these have been identified for each fluid. The Co based MRFs have the highest resistance to viscous transition among the different MRFs. The magnetic stimulus aids the retaining of the elastic nature up to higher strains. Further, the Mn based MRFs also possess the highest stress relaxation caliber, which is further enhanced in the presence of field and higher strains. The study shows that the Mn ferrite based MRFs are the most suitable when the utility requires higher yield stress. However, when applications require the micro-structural integrity of the MRF to be retained in dynamic conditions, Co ferrite based MRFs are more robust, due to their improved magneto-viscoelastic behavior. It is thus shown that introduction of a suitable doping atom in the $M^{2+}$ location of $MFe_2O_4$ structure can definitely improve the magnetorheological and magneto-elastoviscous performances of the corresponding MRFs compared to simple $Fe_3O_4$ based MRFs. The findings may have important implications in efficient design and development of nano-MRFs for variant applications and utilities.

## Supplementary Material

The supplementary material document contains additional information, data, tables and plots of the complex fluid characterization, rheological behavior and additional data on the magneto-elastoviscous response of the fluids.

## Conflict of Interest

The authors declare having no conflicts of interest with any individual or agency with respect to this article.

## Acknowledgements

PD thanks the Department of Mechanical Engineering, IIT Ropar for financial support towards the present work. Also partial funding through the Interdisciplinary project (CDT) by IIT Ropar is acknowledged. AC would like to thank Ministry of Human Resource Development, Govt. of India for the Ph.D. scholarship at IIT Ropar.